\documentstyle[12pt]{article}
\begin{document}
\setlength{\unitlength}{0.7mm}\thicklines
\begin{flushright}
IHEP 94-46\\
8 April 1994
\end{flushright}

\begin{center}
{\bf Model for anomalous $\Upsilon'' \to \Upsilon \pi \pi$ transition\\
with account of chiral phase shift \\ in hadron bag of quarkonium}
\end{center}

\vspace*{0.5cm}
\begin{center}
V.V.Kiselev\footnote{E-mail address: kiselev@mx.ihep.su}, A.K.Likhoded\\
{\it Institute for High Energy Physics,\\
Protvino, Moscow Region, 142284, Russia.}
\end{center}

\begin{abstract}
We assume that a rather large bag of the $\Upsilon''(3S b\bar b)$
heavy quarkonium is in the phase of spontaneously broken chiral symmetry,
in contrast to the $\Upsilon(1S b \bar b)$ hadron bag, having small size
and being in the phase of exact chiral symmetry, so that the shift of the
phases in the  $\Upsilon'' \to \Upsilon \pi \pi$ transition is the additional
source for the $\pi\pi$ pairs. This source causes an anomalous
$\pi\pi$ invariant mass distribution, which is experimentally observed.
\end{abstract}

\section{Introduction}

Description of the quark and gluon hadronization or their confinement
is one of the most important, but complex problems in QCD, because of the
nonperturbative character of the cosinderation in the region of large
distances ($r^{-1} \sim \Lambda_{QCD}$), where hadrons are forming.

However, there is a set of some circumstances, at whose presence the problem
is simplified. So, for instance, considering hadrons,
containing a single heavy quark $Q$ ($m_Q \gg \Lambda_{QCD}$), one succeded
to state some universal regularities, characterizing the bound states of
quarks. Neglecting corrections over the low ratio $\Lambda_{QCD}/m_Q$, one
can consider the heavy quark inside meson to be a static source of the gluon
field. In that case, effective lagrangian of heavy quarks seems to be
symmetric with respect to the substitution of a heavy quark with velocity
$\vec{v}$ by any other heavy quark, moving with the same velocity and having an
arbitrary orientation of its spin. This effective heavy quark symmetry
\cite{1} allows one to state the universal characteristics independent of the
heavy quark flavour: a scaling law for leptonic constants of heavy mesons,
containing a single heavy quark, a behaviour of form factors for semileptonic
transitions between heavy hadrons with a single heavy quark, so that the
behaviour is determined by the only universal function with a fixed
normalization at the border of the phase space, when the hadrons are at rest
with respect to each other.

In the case of heavy quarkonium, composed by heavy quark and heavy antiquark
($Q\bar Q$), nonrelativistic motion of the quarks causes the development
of phenomenological potential models, which rather accurately discribe the mass
spectra of the charmonium and bottomonium \cite{2} and give qualitative
picture of the quark interactions, when the Cou\-l\-omb - like interaction
at small distances is changed by the linearly rising confining potential at
large distances. A low value of the ratio $\Lambda_{QCD}/m_Q$ allows one,
rather reliably and with a success, to apply the QCD sum rules \cite{3}
to the description of the heavy quarkonium properties. The QCD sum rules
combine the perturbative calculations with the extraction of a contribution by
vacuum expectation values of composite operators. Recently in the framework
of the QCD sum rules, scaling relations for leptonic constants of heavy
quarkonia \cite{4} and for coupling constants of
the heavy quarkonia with heavy
mesons ($g_{\Upsilon(4S) B\bar B}$ and $g_{\psi(3770) D\bar D}$) \cite{5}
have been derived. Those laws are in a good agreement with the experimental
data \cite{5a}.

Consideration of hadron transitions between levels of
the charmonium and bottomonium families leads to some universal properties,
too. First of all, in the case of nonrelativistically moving heavy
quarks, bound into colour-singlet state,
one has stated the multipole expansion for the gluon
emission in QCD \cite{6}. This expansion has been stood
as the basis for the description of the hadron transitions between the
heavy quarkonium levels in the framework of QCD \cite{7,8,9}. One has
introduced the assumption on the factorization of the transition matrix
element, which can be expressed in the form of the product of the amplitude
for the multipole gluon emission by the heavy quarks and the amplitude
for the gluon hadronization, i.e. one has assumed that the gluon conversion
into hadrons does not depend on the heavy quarks. As it has been shown in
refs.\cite{8,9}, the calculation of the gluon conversion into hadrons admits a
nonperturbative description in the framework of the low-energy theorems in
QCD. Thus, one has stated the universal regularity for the hadron transitions
between the heavy quarkonium levels. So, for example, in the case of emission
of massless $\pi$ mesons in the transition between the S-wave vector states
of the heavy quarkonium ($Q\bar Q$), the matrix element has the form
\begin{equation}
M(\bar n ^3S_1 \to n ^3S_1 + \pi^+ \pi^-) =
A^Q(\bar n,n)\;(\bar \epsilon^\mu \cdot \epsilon_\mu)\;\frac{2\pi}{b}\;q^2\;,
\label{1}
\end{equation}
where $q$ is the $\pi\pi$ pair momentum, $\bar \epsilon$, $\epsilon$ are the
polarization vectors of the heavy quarkonia,  $b=11- 2n_f/3$, $n_f=3$ is
the number of light quarks, and $A^Q(\bar n,n)$ is calculated over
wave functions of the heavy quarkonia \cite{7}. One can see from eq.(\ref{1}),
that the $\pi\pi$ pair is in the S-wave. As it has been shown in ref.\cite{9},
eq.(\ref{1}) is correct with the accuracy by very small correction, which
includes the D-wave of the $\pi\pi$ pair, and corrections due to nonzero mass
of the $\pi$ meson are inessential and expressed in the form of the shift
of $q^2$ to $q^2+\delta^2$, where $\delta \sim m_\pi$.

As one can see from figs.1  ,b, expression (\ref{1}) is in a good
agreement with the experimental data on the transitions of
$\psi' \to \psi \pi\pi$ \cite{10} and
$\Upsilon' \to \Upsilon \pi \pi$ \cite{11}. It agrees with the consideration,
early performed in the framework of soft pion technique and PCAC \cite{12}.

The later data by CLEO \cite{13} were in the agreement  with the results
of ref.\cite{11} on the $\Upsilon' \to \Upsilon \pi^+ \pi^-$
transition, but at the first time the CLEO has observed the invariant mass
spectrum of the $\pi\pi$ pairs in the decay of
$\Upsilon'' \to \Upsilon \pi^+ \pi^-$, so that the spectrum is in explicit
contradiction with the result (\ref{1}) (see fig.2).

The difference between the mass spectra of the $\pi$ meson pairs from the
decays of $\Upsilon' \to \Upsilon \pi^+ \pi^-$ and
$\Upsilon'' \to \Upsilon \pi^+ \pi^-$ can be explained by the presence
of additional source of the pions in the case of the $\Upsilon''$ state.

In papers of ref.\cite{14} the assumption has been made on the presence of
hybrid isovector $\Upsilon_1(b\bar b q\bar q)$ state, having the mass
close to the $\Upsilon''$ mass and strongly coupled to the ordinary
$\Upsilon$-particles due to the $\pi$ meson emission. Then the mechanism
of the cascade $\Upsilon'' \to \Upsilon_1 \pi \to \Upsilon \pi\pi$ decay
would be dominant and giving an essential rising of the contribution
by small invariant masses of the $\pi\pi$ pair \cite{14}. However, in
this case the $\Upsilon_1$ state must be observed in the $\Upsilon(4S)$
decays with a rather large probability ($\sim 1\%$), which is not discovered
empirically. Moreover, the experimental data on the $\Upsilon \pi$ mass
spectrum is in a deep contradiction with the distribution, expected at the
$\Upsilon_1$-resonance presence \cite{14,15}.

As it has been shown in refs.\cite{17,18}, an essential contribution into
the region of the low invariant masses of the $\pi\pi$ pair might give an
underthreshold production of the $B\bar B$ pairs, emitting the $\pi$
mesons. In that case, an additional complex parameter appears, so that it
determines the relative contribution by the coupled $B\bar B$ pairs with
respect to the standard multipole emission of gluons, converted into
the $\pi$ mesons.  Fitting the experimental data on the $\pi\pi$ pair mass
spectrum in the $\Upsilon'' \to \Upsilon \pi^+ \pi^-$ decay shows that the
additional contribution must be comparable with the standard one \cite{16}.
However, the coupled channel contribution must cause some characteristic
angle correlations in the decay of $\Upsilon'' \to \Upsilon \pi^+ \pi^-$,
but those correlations are not observed \cite{15,16}.
Furthermore, it has been recently shown that the consistent description of
the coupled $B\bar B$ channel contribution, accounting the $B\bar B$
influence on some other characteristics of the $\Upsilon$-particle family
(mass spectrum, leptonic widths) \cite{19}, leads to the change of the
$\pi\pi$ mass spectrum in the low mass region. However, the change is
very less than the value, which is observed experimentally \cite{20}.

Thus, all early offered models for the anomalous
$\Upsilon''\to\Upsilon\pi^+\pi^-$ transition are empirically rejected.

In the present paper we propose the model for the anomalous distribution
over the $\pi\pi$ pair mass in the decay of
$\Upsilon'' \to \Upsilon \pi^+ \pi^-$ with taking into account the additional
source of the pion production due to a phase transition. We assume that
the phase transition is caused by the following reasons.

The heavy quarkonium system is the only quark system, whose sizes are varied
in a rather broad limits, since, in contrast to bound states of light quarks,
the scale of distances inside the heavy quarkonium is determined not only by
the confinement energy through the potential parameters, it is also determined
by the heavy quark
masses. Namely, the size of the heavy quarkonium hadron bag is determined
by the heavy quark motion. So, for instance, in the framework of the
nonrelativistic potential models for the vector $\Upsilon$-particle
states  one has \cite{2}
\begin{eqnarray}
r(\Upsilon) & \simeq & 0.3\;fm\;, \nonumber \\
r(\Upsilon') & \simeq & 0.5\;fm\;,  \\
r(\Upsilon'') & \simeq & 0.8\;fm\;, \nonumber
\end{eqnarray}
so that the sizes of the 1S- and 2S-levels are majorly determined by the
Coulomb-like part of the potential, and the 3S-level size is already
essentially determined by the lenearly rising confining part of the potential.
Thus, phenomena, taking place at the large distances and, hence, connected to
the confinement, must be valuably displayed in the properties of the
$\Upsilon''$ state. One of the characteristic phenomena of the confinement
is the generation of quark codensate and, therefore, the appearance of
effective, constituent mass of light quarks, moving in the condensate
medium. This phenomenon finds adequate description in the framework of
spontaneous chiral symmetry breaking \cite{21}, which has the form of the
phase transition.

In our furthercoming calculations we assume that the hadron bag of the
$\Upsilon''$ state is rather large for the quark-gluon condensate influence
to be observable, i.e. the influence by the phase of the spontaneously
broken chiral symmetry is valuable. At the moment, the $\Upsilon'$ and
$\Upsilon$ sizes are small and their hadronic bags are in the phase of
the exact chiral symmetry with a minimal influence of the surrounding
medium. Thus, the $\Upsilon'' \to \Upsilon \pi^+ \pi^-$ decay must be
accomponied by the phase transition, corresponding to the different conditions
for the existence of the $\Upsilon''$ and $\Upsilon$ hadron bags (see fig.3).

As we have pointed, the hadronic transitions between the levels of the
heavy quarkonium are the only circumstances, where we might, in principle,
explicitly observe the shift of the chiral symmetry phases. In all other cases
we deal with a situation, when the exact chiral symmetry phase exists only
virtually during a rather short time, and further it evolves to the phase of
spontaneous chiral symmetry breaking (the situation is like  hard
production of parton jets with the further fragmentation into hadrons) .

The spontaneous process of the phase transition means the energy release,
which shows to be the additional source of the $\pi$ meson pair
production with the anomalous distribution over the invariant mass of the pair,
when the low values of the mass are nor suppresed.

In section 2 we show that with account of the phase transition, the
soft pion approximation leads to the following expression for the
matrix element of the $\Upsilon'' \to \Upsilon \pi^+ \pi^-$ decay
\begin{equation}
M(\Upsilon'' \to \Upsilon \pi^+ \pi^-) = A^b(3S,1S)\;
\epsilon''_\mu\cdot \epsilon^\mu\;\frac{2\pi}{b}\;(q^2-\mu_1^2+i \mu_2^2)\;,
\label{3}
\end{equation}
so that
\begin{equation}
\mu_1^2-i \mu_2^2 = \frac{2}{f_\pi^2}\;e^{i\alpha_\phi}\;m_{const}\;
<0|(\bar u u +\bar d d)|0>\;, \label{4}
\end{equation}
where $m_{const}$ is the constituent light quark mass,
$\alpha_\phi$ is the difference between the complex phases, characterizing
the states with the spontaneously broken and exact chiral symmetry.
In expression (\ref{4})  the quark mass and condensate values are taken
at the scale, corresponding to the size of the decaying state, so that
expressions (\ref{3}), (\ref{4}) convert into eq.(\ref{1}) if one accounts
that eq.(\ref{1}) is valid for the states, having the sizes
less than the critical scale, i.e. at $\alpha_\phi = 0$, $m_{const}=0$.
We make also the phenomenoligical analysis of the values in
eqs.(\ref{3}), (\ref{4}) and show, that eqs.(\ref{3}), (\ref{4}) are
in agreement with the experimental data at some reasonable values of the
parameters.

In section 3 we show that with account of the contribution by the additional
source, connected with the phase transition, eqs.(\ref{3}), (\ref{4})
do not contradict with the Adler theorem \cite{22}. We modify the analysis,
made in the framework of the soft pion technique and PCAC \cite{12}.

In section 4 the obtained results are discussed.

\section{QCD model for the $\Upsilon'' \to \Upsilon \pi^+ \pi^-$ transition}

In the framework of QCD, the model for hadronic transitions between levels
of the heavy quarkonium family is based on the multipole expansion
for the gluon emission by the heavy nonrelativistic quarks with the
postcoming gluon hadronization, which does not depend on the heavy
quark motion.

\subsection{Dipole emission}

Action, corresponding to the coupling of the heavy quarks to the gluon field,
generally has the form
\begin{equation}
S_{int} = -g\;\int d^4x\;A_\mu^a(x)\cdot j^\mu_a(x)\;, \label{5}
\end{equation}
where $g$ is the coupling constant of the quarks to the gluons.

In the case of gluon emission by the heavy nonrelativistic quark and
antiquark, being in the colour-singlet vector nS-state, in accordance
with eq.(\ref{5}) one can write down
\begin{equation}
S_{int} = -g\;\int dt d^3\vec{x}\;A_\mu^a(t,\vec{x})\cdot
(j^\mu_a(t,\vec{x}+\vec{r}/2)+\bar j^\mu_a(t,\vec{x}-\vec{r}/2))\;
\Phi(\vec{r}) d^3\vec{r}\;, \label{6}
\end{equation}
where $j$ and $\bar j$ are the quark and antiquark currents, which in the
leading approximation have the form
\begin{eqnarray}
j^\mu_a(t,\vec{x}+\vec{r}/2) & = &
(j^0_a(t,\vec{x}+\vec{r}/2),\vec{0})\;, \nonumber \\
\bar j^\mu_a(t,\vec{x}-\vec{r}/2) & = &
(-j^0_a(t,\vec{x}-\vec{r}/2),\vec{0})\;,\\
j^0_a(t,\vec{r}) & = & \delta(\vec{r})\;\frac{\lambda_a^{ij}}{2}\;, \nonumber
\end{eqnarray}
where $\lambda_a^{ij}$ are the Gell-Mann matrices. In eq.(\ref{6}) the
$\Phi(\vec{r})$ function corresponds to the amplitude to find the heavy quarks
at the distance $\vec{r}$ between the quarks, so that
\begin{equation}
\Phi(\vec{r}) = \frac{1}{\sqrt{3}}\delta^{im} \Psi_n(\vec{r})\;
\Psi_f^{mj}(\vec{r})\;K(s_n,f)\;, \label{8}
\end{equation}
where $\Psi_n(\vec{r})\delta^{im}/\sqrt{3}$ is the heavy quarkonium wave
function, and $\Psi_f^{mj}(\vec{r})$ is the wave function of the colour-octet
state of the heavy quark and antiquark system, produced after the single
gluon emission. The factor $K(s_n,f)$ respects to the spin coefficient
for initial and final states in the emission (note, in the leading
approximation (see ref.\cite{1}) the heavy quark spin is decoupled from the
interaction with the gluons, so that the spin state of the quarkonium
is not changed).

Thus, at low $\vec{r}$ values,
expression (\ref{6}) for the gluon interaction with the heavy quarkonium
can be rewritten in the form
\begin{eqnarray}
S_{int} & = & -g\;\int dt d^3\vec{x}\;A_0^a(t,\vec{x})\cdot
r^k \partial_k\delta(\vec{x})\;\frac{\lambda_a^{ij}}{2} \Psi_n(\vec{r})
\Psi_f^{ji}(\vec{r}) \;K(s_n,f) \;d^3\vec{r}\; \nonumber \\
{}~ & = & g\;\int dt \;
r^k \partial_k A_0^a(t,\vec{x})\;\frac{\lambda_a^{ij}}{2} \Psi_n(\vec{r})
\Psi_f^{ji}(\vec{r}) \;K(s_n,f) \;d^3\vec{r}\;.
\end{eqnarray}
One can see, that with the accuracy by the higher corrections over both
the quark motion inside the quarkonium and $g$,
the hamiltonian of the heavy vector quarkonium interaction
with the gluon field has the form
\begin{equation}
H_{int} = -g \;\vec{r}\cdot \vec{E^a}\;,
\end{equation}
i.e. the hamiltonian has the form of the dipole chromoelectric interaction
($\vec{E^a} \simeq \vec{\partial}A_0^a+...$).

The strict derivation of the formulae for the multipole emission in QCD
is performed in papers of ref.\cite{6}.

Then in the leading approximation the matrix element for the
$\bar n^3S_1 \to n^3S_1 +gg$ transition can be rewritten in the form
\begin{eqnarray}
M(\bar n^3S_1 \to n^3S_1+gg) & = & 4\pi \alpha_S\;E_k^a E_m^b\;
\cdot \nonumber \\
{}~ & ~ & \int d^3r d^3r'\;r_k r'_m\;G^{ab}_{s_{\bar n},s_n}(r,r')\;
\Psi_{\bar n}(r) \Psi_n(r'), \label{11}
\end{eqnarray}
where $G^{ab}_{s_{\bar n},s_n}(r,r')$ corresponds to the propagator of the
colour-octet state of the heavy quarkonium
\begin{equation}
G = \frac{1}{\epsilon - H_{Q\bar Q}^c}\;,
\end{equation}
where $H_{Q\bar Q}^c$ is the hamiltonian of the coloured state.

The consistent calculation of the integral in eq.(\ref{11})
is performed in papers of ref.\cite{7}, where it has been shown that
\begin{equation}
M(\bar n^3S_1 \to n^3S_1+gg) = \alpha_S\;E_k^a E_k^a\;
A^Q(\bar n, n)\;\bar \epsilon_\mu\cdot \epsilon^\mu\;, \label{13}
\end{equation}
so that the information about the heavy quark motion mainly contains in the
$A^Q(\bar n, n)$ factor.

Generally, the factorization of the heavy quark motion can be written in
the form
\begin{equation}
<\bar n^3S_1|T|n^3S_1\pi\pi> = <0|O_\phi\;\alpha_S\;E^2|\pi\pi>\;
A^Q(\bar n, n)\;\bar \epsilon_\mu\cdot \epsilon^\mu\;, \label{14}
\end{equation}
where, as it has been discussed above, the operator
$O_\phi = |0><0_\phi| = 1+\phi C_\phi$ corresponds to the phase transition
between the $\bar n$ and $n$ states and depends on the order parameter
$\phi$, determined by the quark codensate and the constituent light quark mass,
so that the wave function of the initial state can be presented in the form
\begin{equation}
<\vec{r}|\bar n^3S_1> = \Psi_n(\vec{r})|0_\phi>\;,
\end{equation}
where $|0_\phi>$ is the vacuum state, accounting the perturbance of the quark
condensate (see fig.3), so that, acting on the $|0_\phi>$ state, the
operators of the heavy quark creation produce the real state of the heavy
quarkonium with the account of the phenomena of the spontaneous chiral symmetry
breaking.
Indeed, the operators of the heavy quark creation
produce the heavy quarks with
the account of a dispersion law, which takes into account the presence of the
condensate at the large distances. However, these operators do not produce
the perturbances of the condensate state (see fig.3),
and this change in the vicinity of the quarkonium is produced by
the $O_\phi$ operator, depending on the hadronic bag size of the quarkonium.
In the case of the heavy quarkonium with the small size ($\Upsilon$,
$\Upsilon'$), the quarks have rather large virtualities to have no influence
on the condensate state, so that $O_\phi(\Upsilon, \Upsilon') \equiv 1$.
In the case of the $\Upsilon''$ state, whose size is so large, that the
hadron bag essentially depends  on the condensates, the operator
$O_\phi(\Upsilon'')\ne 1$ and it contains the chiral noninvariant term
$\phi C_\phi$, $[Q_5,C_\phi] \ne 0$, so that the chiral symmetry is
broken spontaneously, i.e. due to the noninvariance of the $|0_\phi>$ state
at the chiral invariant action of QCD $S_{QCD}= \int d^4x L_{QCD}(x)$.

Note, eq.(\ref{14}) means that the factorization of the heavy quark
contribution from the contribution by the large distances is kept valid
in the weaker form. Namely, the amplitude of the gluon conversion into the
$\pi$ mesons does not contain the operators of the heavy quark creation and
annihilation, and, hence, the heavy quark motion is factorized. However,
the amplitude contains the transition operator, connnected to the changes
of the hadron bag of the heavy quarkonium.

Thus, the calculation of the matrix element for the hadronic transition
$\bar n^3S_1 \to n^3S_1 + \pi \pi$ is reduced to the problem of the
description for  the gluon conversion into the $\pi$ mesons
\begin{equation}
\bar M = <0|O_\phi\;\alpha_S\;E^2|\pi\pi>\;.\label{16}
\end{equation}

\subsection{Gluon conversion into $\pi$ mesons with account of
phase transition}

In this section we show that in the case of soft pions the matrix element
(\ref{16}) can be exactly calculated without an application of the
perturbation theory, which can not be reliable for the hadronization at the
large distances.

Following refs.\cite{8,9}, one can get
\begin{eqnarray}
\alpha_S E^2 & = &
\alpha_S\;\frac{E^2+H^2}{2} + \alpha_S\;\frac{E^2-H^2}{2}\nonumber \\
{}~ & = &
\alpha_S \theta^g_{00} + \frac{2\pi}{b} \theta^g_{\mu\mu}(1+O(\alpha_S))\;,
\label{17}
\end{eqnarray}
where
\begin{equation}
\theta^g_{\mu\mu} = -\; \frac{\beta(\alpha_S)}{4\alpha_S}\; G^a_{\mu\nu}
G_a^{\mu\nu}
\end{equation}
is the tensor of the gluon energy-momentum,
$$
\beta(\alpha_S) = -\;\frac{b}{2\pi}\;\alpha_S^2
$$
is the Gell-Mann -- Low function in QCD, $b=11-2n_f/3$, $n_f=3$.

As it has been shown in ref.\cite{9}, the first term in the right hand side
of eq.(\ref{17}) has a low value, which will be neglected in what follows,
so that
\begin{equation}
\bar M = <0|O_\phi\;\theta^g_{\mu\mu}|\pi\pi>\;.\label{18}
\end{equation}
Neglecting the current mass of the light quarks, one can write down
\begin{equation}
\theta^g_{\mu\mu} = \theta^{QCD}_{\mu\mu}\;.\label{19}
\end{equation}
Further, by the law of the energy-momentum conservation one has
\begin{equation}
<0_\phi|\pi^+(q_1)\pi^-(q_2)> = 0,\;\;\;\;(q_1+q_2)^2 > 0\;.
\end{equation}
Therefore the matrix element (\ref{18}) can be rewritten in the form
\begin{equation}
\bar M = \frac{2\pi}{b}\;<0|(\theta^{QCD}_{\mu\mu}+\phi
[C_\phi,\theta^{QCD}_{\mu\mu}]\;)|\pi^+\pi^->\;,
\end{equation}
so that the term, neglected in refs.\cite{8,9}, has the form
\begin{equation}
\phi \;[C_\phi,\theta^{QCD}_{\mu\mu}] = \Delta \theta^{QCD}_{\mu\mu}(\phi)\;,
\end{equation}
where $\Delta \theta^{QCD}_{\mu\mu}(\phi)$ corresponds to the effective
contribution to the trace of the energy-momentum tensor in the presence of
the source, the order parameter, breaking the chiral symmetry and depending
on the scale. One can easily see, the order parameter for the chiral
symmetry breaking is the constituent light quark mass, related with the nonzero
quark condensate, so one gets
\begin{equation}
\Delta \theta^{QCD}_{\mu\mu}(\phi) = e^{i\alpha_\phi}\;m_{const}(r_{\bar n})\;
(\bar u u + \bar d d)\;, \label{23}
\end{equation}
where $\alpha_\phi$ is the difference of the complex phases for the
$|0_\phi>$ and $|0>$ states, and the value of the constituent mass is
determined by the size of the decaying $\bar n S$ quarkonium.

{}From eqs.(\ref{18})-(\ref{23}) it follows that
\begin{equation}
\bar M = \frac{2\pi}{b}\;(<0|\theta^{QCD}_{\mu\mu}|\pi^+\pi^-> +
e^{i\alpha_\phi}\;m_{const}(r_{\bar n})\;
<0|(\bar u u + \bar d d)|\pi^+\pi^->)\;,
\label{24}
\end{equation}
so that if the size of the decaying quarkonium is rather small, one has
$\alpha_\phi = 0$ and $m_{const} = 0$.

As it has been shown in ref.\cite{9}, the matrix element
\begin{eqnarray}
<0| \theta^{QCD}_{\mu\nu} |\pi^+(q_1) \pi^-(q_2)> & = &
A r_\mu r_\nu + B q^2 g_{\mu\nu} + C q_\mu q_\nu\;, \label{25}\\
r & = & q_1 - q_2\;, \nonumber\\
q & = & q_1 + q_2\;, \nonumber
\end{eqnarray}
can be calculated by the use of three following conditions.

1) The law of energy conservation
\begin{equation}
q^\mu\;\theta^{QCD}_{\mu\nu} = 0\;.\label{26}
\end{equation}

2) The chiral invariance of the $\theta^{QCD}_{\mu\nu}$ value, which means
that
\begin{equation}
<0| \theta^{QCD}_{\mu\nu} |\pi^+(q_1) \pi^-(q_2)>_{|q_1 \to 0} =
\frac{i}{f_\pi}\; <0|\;[\theta^{QCD}_{\mu\nu}, Q_5^+]\;|\pi^-(q_2)> = 0\;.
\label{27}
\end{equation}

3) The normalization of the energy-momentum tensor
\begin{equation}
<0| \theta^{QCD}_{\mu\nu} |\pi^+(p) \pi^-(p)> = 2 p_\mu p_\nu\;. \label{28}
\end{equation}

{}From eqs.(\ref{25})-(\ref{28}) it follows that
\begin{equation}
<0| \theta^{QCD}_{\mu\nu} |\pi^+(q_1) \pi^-(q_2)>  =
\frac{1}{2}\;( r_\mu r_\nu +  q^2 g_{\mu\nu} - q_\mu q_\nu)\;, \label{29}
\end{equation}
and
\begin{equation}
<0| \theta^{QCD}_{\mu\mu} |\pi^+(q_1) \pi^-(q_2)>  = q^2\;, \label{30}
\end{equation}
Further, one can easily find for soft pions (see (\ref{27})) that
\begin{equation}
<0|\;(\bar u u + \bar d d)\;|\pi^+(q_1)\pi^-(q_2)> =
-\;\frac{2}{f^2_\pi}\;<0|\;(\bar u u + \bar d d)\;|0>\;, \;\;\;q_{1,2} \to 0\;.
\label{31}
\end{equation}
Then for the gluon conversion into the $\pi$ meson pair one gets the expression
\begin{equation}
\bar M = \frac{2\pi}{b}\;(q^2 - \frac{2}{f^2_\pi}\;
e^{i\alpha_\phi}\;m_{const}(r_{\bar n})\;<0|(\bar u u + \bar d d)|0>)\;,
\label{32}
\end{equation}
where $f_\pi \approx 132$ MeV. From eqs.(\ref{14}), (\ref{16}), (\ref{18}) and
(\ref{32}) it follows that the matrix element of the
$\Upsilon'' \to \Upsilon \pi^+ \pi^-$ transition has the form
(\ref{3}), (\ref{4}).

\subsection{Phenomenological analysis of the
$\Upsilon''\to\Upsilon\pi^+\pi^-$ transition}

The differential distribution of the decay width of
$\Upsilon'' \to \Upsilon \pi^+ \pi^-$ over the invariant mass
$m_{\pi\pi}$ ($q^2 = m^2_{\pi\pi}$) of the $\pi\pi$ pair has the form
\begin{equation}
\frac{d\Gamma}{d m_{\pi\pi}} = \bar A\;|\vec{k}_{\pi\pi}|\;
|\vec{q}_\pi|\;((q^2-\mu_1^2)^2+\mu_2^4)\;,\label{33}
\end{equation}
where $\bar A$ is an effective constant, which can be defined by the explicit
form of the matrix element (\ref{3}), (\ref{4}), $\vec{q}_\pi$ is the
$\pi$ meson momentum in the $\pi\pi$ pair rest frame,
\begin{equation}
|\vec{q}_\pi| = \frac{1}{2}\;\sqrt{q^2-4m_\pi^2}\;,
\end{equation}
$\vec{k}_{\pi\pi}$ is the $\pi$ meson pair momentum in the rest frame of
the decaying quarkonium.

A rather good agreement of eq.(\ref{33}) with the experimental data
(see fig.2)  is achieved at the following values of the paprameters
\begin{eqnarray}
\mu_1 & \simeq & 680\;\;MeV\;, \label{35}\\
\mu_2 & \simeq & 400\;\; MeV\;,\label{36}
\end{eqnarray}
which corresponds, in agreement with eq.(\ref{4}), to the values
\begin{eqnarray}
m_{const}(r_{\Upsilon''}) & \simeq & 140\;\; MeV\;, \label{37}\\
cos \alpha_\phi & \simeq & -0.95\;,\;\;\;|\pi\pm\alpha_\phi| \simeq 19^\circ\;,
\label{38}
\end{eqnarray}
if one supposes that at the scale of the order of
$r(\Upsilon'') \simeq 0.8$ fm \cite{3}, one has
\begin{equation}
<0|\bar u u|0> = <0|\bar d d|0> \simeq -(250\;\;MeV)^3\;. \label{39}
\end{equation}
As one can see from eq.(\ref{37}), the value of the constituent light
quark mass, which is displayed in the anomalous
$\Upsilon'' \to \Upsilon \pi^+ \pi^-$ transition, is slightly less than the
mass of the valent light quark in ordinary hadrons.
It is naturally, since the constituent mass is growing with the rising of the
hadron size, and in the case of the heavy $\Upsilon''$ quarkonium,
whose size is less than the sizes of the ordinary hadrons with the valent
light quarks, the manifested mass is less than it is in the ordinary hadrons
\footnote{At the choice of the less value of the light quark condensate,
whose value  depends on the scale (for example,
$<0|\bar u u|0> = <0|\bar d d|0> \simeq -(200\;\;MeV)^3$ ), the estimate
of the constituent light quark mass will give the greater value than in
eq.(\ref{37}), so $m_{const}(r_{\Upsilon''})  \simeq  274\;\; MeV$.}.
Nevertheless, the scale of the value in eq.(\ref{37}) can be considered
as reasonable, since $m_{const} \sim \Lambda_{QCD}$.

As for the complex phase difference ($\alpha_\phi$) between the states
with the exact and spontaneously broken chiral symmetry, it is considered as
the external parameter of the present model.

Thus, the modified matrix element of the
$\Upsilon''\to \Upsilon\pi^+\pi^-$ transition with the account of the shift
of the chiral symmetry phase in the quarkonium hadronic bag, is in the
reasonable agreement with the experimental data on the $\pi\pi$ pair
mass spectrum.

\section{$\Upsilon'' \to \Upsilon \pi^+ \pi^-$ transition in soft
pion technique}

The consideration of the hadronic transitions between the S-wave levels
of the heavy quarkonium has been performed in the framework of PCAC in
papers of ref.\cite{12} and it must be modified to account the phase
transition of the quarkonium bag in the case of
$\Upsilon'' \to \Upsilon \pi^+ \pi^-$.

The selfconsistency condition by Adler \cite{22} stands that the amplitude
of the transition of the $b$ state into the $a\pi$ state with the pion
is equal to
\begin{equation}
T(b \to a\pi) = \frac{i}{f_\pi}\;q^\mu\cdot T_\mu\;,\;\;q \to 0\;,
\label{40}
\end{equation}
where $q$ is the $\pi$ meson momentum,
\begin{equation}
T_\mu = <b|\;\bar A_\mu\;|a>\;,\label{41}
\end{equation}
where $\bar A_\mu$ is the axial current without the contribution by
the $\pi$ meson pole.

Then, taking into the account the factorization of the heavy quarkonium wave
functions, one can write down
\begin{equation}
M(\Upsilon'' \to \Upsilon \pi^+ \pi^-) = A\;\epsilon''_\mu\cdot
\epsilon^\mu\;<0_\phi|\;T\;|\pi^+(q_1)\pi^-(q_2)>\;, \label{42}
\end{equation}
so that at $q_{1,2} \to 0$ one gets
\begin{equation}
<0_\phi|\;T\;|\pi^+(q_1)\pi^-(q_2)> = -\;\frac{1}{f^2_\pi}\;
q_1^\alpha q_2^\beta\;<0_\phi|\;T\;\bar A^+_\alpha (q_1) \bar A^-_\beta (q_2)\;
|0>\;. \label{43}
\end{equation}
In $\sigma$ model \cite{21}, which has a rather general character, one
can find that the nonpole contribution into the axial current is equal to
\begin{equation}
\bar A_\alpha^k(x) = -\pi^k(x) \partial_\alpha \sigma(x)\;,
\end{equation}
Then one obtains, that
\begin{eqnarray}
T_{\alpha\beta} & = & <0_\phi|\;T\;\bar A^+_\alpha (q_1) \bar A^-_\beta (q_2)\;
|0> \nonumber \\
{}~ & = & \int d^4x d^4y e^{iq_1x+iq_2y} <0_\phi|\;T\;
\partial^x_\alpha \sigma(x) \partial^y_\beta \sigma(y)\;
\pi^+(x) \pi^-(y)\;|0>. \label{45}
\end{eqnarray}
In the case  of the phase transition, the $\sigma$ field state is changed,
and the matrix element $T_{\alpha\beta}$ gets the coherent contribution,
depending on the order parameter $\phi$ ($\phi \sim f_\pi,\;\;m_{const}$), and
\begin{equation}
\Delta T_{\alpha\beta} = <0|\;\Delta_\phi\;
(T\;\partial^x_\alpha \sigma(x) \partial^y_\beta \sigma(y))\;
\pi^+(x) \pi^-(y)\;|0>\;, \label{46}
\end{equation}
where
\begin{equation}
<0|\;\Delta_\phi\;(T\;\partial^x_\alpha \sigma(x) \partial^y_\beta \sigma(y))
= \int \frac{d^4k}{(2\pi)^4}\;e^{ik(x+y)}\;k_\alpha k_\beta\;
\frac{D_\phi(\mu^2)d\mu^2}{k^2-\mu^2}\;<0| + h.c. \label{47}
\end{equation}
Eq.(\ref{47}) has the form of the coherent emission, caused by the phase
transition. Since there is the massless goldstone $\pi$ meson at the
spontaneous chiral symmetry breaking, the spectral density of the emission
$D_\phi(\mu^2)$ must contain the contribution by the massless particle, so that
\begin{equation}
\Delta T_{\alpha\beta}^\phi(x,y) = <0|\pi^+(x)\pi^-(y)|0>\;\int
e^{ik(x+y)}\;A_\phi\;\frac{k_\alpha k_\beta}{k^2}\;\frac{d^4k}{(2\pi)^4}
+reg.part \label{48}
\end{equation}
where $A_\phi$ is proportional to the order parameter $\phi$.

{}From eq.(\ref{48}) it follows that
\begin{equation}
\Delta T_{\alpha\beta}^\phi(q_1,q_2) = -2 A_\phi\;\frac{1}{q^2}\;
\frac{q_\alpha q_\beta}{q^2} + reg.part \label{49}
\end{equation}
{}From eqs.(\ref{42})-(\ref{49}) one finds that for soft pions
\begin{equation}
M(\Upsilon'' \to \Upsilon \pi^+ \pi^-) = A\;\epsilon''_\mu\cdot
\epsilon^\mu\;\frac{1}{f^2_\pi}\;(A_\phi/2 + B q^2)\;, \label{50}
\end{equation}
where the complex number $A_\phi$ is determined by the constituent light quark
mass, which is displayed at the scale of the hadronic bag of the $\Upsilon''$
state. The value of the $A_\phi$ to $B$ ratio can be theoretically defined
only in the framework of detailed quark model for the axial currents, as it
has been performed in section 2.

Thus, taking into the account the phase transition, the matrix element
of the axial currents (\ref{45}) gets the singular term (\ref{49}),
connected to the constituent light quark mass, which is not equal to zero
for rather large hadron bag of the $\Upsilon''$ level. Hence,
$T_{\alpha\beta}$ is not regular at $q_{1,2} \to 0$, so this regularity
condition is necessary for the applicability of the Adler theorem \cite{22},
that stands that, at the absence of singularities, the amplitude of the
$\pi$ meson emission with the momentum $q\to 0$ must tend to zero.
Eq.(\ref{50}) modifies the cosideration, early performed in the framework
of the same technique of soft pions and PCAC in papers of ref.\cite{12}.

\section{Conclusion}

In the present paper we have offered the model of the anomalous mass
spectrum of the $\pi$ meson pair in the decay of
$\Upsilon'' \to \Upsilon \pi^+ \pi^-$, taking into the account the phase
transition of the heavy quarkonium hadronic bag.

The heavy quarkonium is an exceptional system, since, first, concerning the
hadronic transitions between the quarkonium levels, one can factorize
the heavy quark motion from the processes of the gluon and light quark
hadronization, so that due to the nonrelativistic motion of the heavy quarks
the consideration of the gluon emission by the heavy quarks allows one to
apply the multipole expansion in QCD. Second, the heavy quark motion
defines the hadronic bag size of the quarkonium within the broad limits
from 0.3 to 0.9 fm, so that such critical phenomenon as the spontaneous
chiral symmetry breaking, taking place in the form of the phase transition,
say, at $r \sim 0.7$ fm, can essentially change the dynamics of the hadron
transition for the quarkonia with $r < 0.7$ fm and $r > 0.7$ fm. This dynamical
difference has been shown in the present paper for the large hadronic bag of
the $\Upsilon''$ level with $r(\Upsilon'') \simeq 0.8 $ fm.

As for the model parameters, the fitting the experimental data on the
$\Upsilon'' \to \Upsilon \pi^+ \pi^-$ decay gives the reasonable value of
the constituent light quark mass, which is displayed in the
$\Upsilon'' \to \Upsilon \pi^+ \pi^-$ transition, so that
$m_{const} \simeq 140$ MeV. However, note that for the description of the
$\Upsilon'' \to \Upsilon \pi^+ \pi^-$ transition, having rather large phase
space: $0 < q^2 < 1\;\;GeV^2$, we have supposed the model parameters to be
constant values (the complex phase difference between the state with the
source and the state without the later, the constituent light quark mass),
because the behaviour of the matrix element is generally determined by
the facts, that the amplitude is not equal to zero at  $q^2 \approx 0$ ,
and it rises at $q^2 \approx 1\;\;GeV^2$ as $q^2$,  and the constant and
rising contributions destructively interfere in the centre of the mass
spectrum. Therefore, in the case of the $\Upsilon'' \to \Upsilon \pi^+ \pi^-$
transition the above approximation can be considered as reasonable.
If the phase space of the decay is small, as it takes place in the
$\Upsilon'' \to \Upsilon' \pi^+ \pi^-$ decay, then the behaviour of the
matrix element will be essentially determined by the dependence of the
model parameters on $q^2$,  especially, by the dependence of the
complex phase difference $\alpha_\phi(q^2)$, which, in contrast to
$m_{const}$, must tend to zero at $q^2 = 4m_\pi^2$, for example.
Therefore, the description of the $\pi$ meson pair mass spectrum in the
$\Upsilon'' \to \Upsilon' \pi^+ \pi^-$ transition needs the introduction of
some additional assumptions, which we suppose to discuss elsewhere.

In addition to the $\Upsilon'' \to \Upsilon \pi^+ \pi^-$ ,
$\Upsilon' \to \Upsilon \pi^+ \pi^-$, $\Upsilon'' \to \Upsilon' \pi^+ \pi^-$
transitions, one experimentally observes the $\psi' \to \psi \pi\pi$ decay,
which is well described by the matrix element without the account of the
phase transition, although the $\psi'$ size is the same as the $\Upsilon''$
size: $r(\psi') \simeq 0.8$ fm. We think that the qualitative picture
of the phase transition, which we have presented in the previous sections
must be carefully applied to the heavy charmonium. This is caused by the
following reasons. First, the uncertainty in the $b$-quark mass value is low,
$\Delta m_b/m_b \sim \Lambda_{QCD}/m_b < 10 \%$. Second, the
$b$-quark Compton length is well defined and it is very small
$\lambda_b \simeq 0.04$ fm. Therefore, the average radius of the bottomonium
is rather distinct border of the hadronic bag of the $\bar b b$ system.
In the case of the $c$-quarks the border of the hadronic bag gets an
essential uncertainty, connected to both the uncertainty in the
$c$-quark mass $\Delta m_c/m_c \sim 30 \%$ and its large Compton length
$\lambda_c \simeq 0.15$ fm. Hence, the important threshold characteristics
as the typical distance between the quarks become not exactly defined value
for the charmonium (for example, $r(\psi') - 2\lambda_c \simeq 0.5$
fm). Hence, the border of the charmonium hadronic bag is not distinct, and
it is impossible  reliably to state {\it a priori} whether the phase
transition in the $\psi' \to \psi \pi\pi$ decay takes place or does not.
The fact is stated by the experimental observation of the invariant
mass spectrum of the $\pi$ meson pair, where one finds the agreement
with the consideration with the exact chiral symmetry in $\psi'$.

Thus, the present model can reasonably be in the agreement with the data
on the $\Upsilon'' \to \Upsilon \pi^+ \pi^-$ transition (with
the account of the data on the angle correlations, which correspond to
the isotropic distributions observed experimentally),
and the model is not in a contradiction with the data on the other analogous
transitions.

\newpage
\begin{center}
{\large Figure captions}
\end{center}
\begin{description}
\item{
Fig. 1.} Mass spectrum of the $\pi$ meson pairs in the decays of
 ) $\psi'\to \psi + \pi\pi$ and b) $\Upsilon' \to \Upsilon + \pi\pi$,
$x=(m_{\pi\pi}-2 m_\pi)/(m_i - m_f- 2m_\pi)$, where $m_i$, $m_f$ are
masses of the initial and final states, respectively.
The curve is obtained in the model with the matrix element (\ref{1}).
\item{
Fig. 2.} Mass spectrum of the $\pi$ meson pairs in the decay of
$\Upsilon'' \to \Upsilon + \pi\pi$.
Dashed line is obtained in the model with the matrix element (\ref{1}),
solid line is the present model with eqs.(\ref{3})-(\ref{4}) and
(\ref{35})-(\ref{39}).
\item{
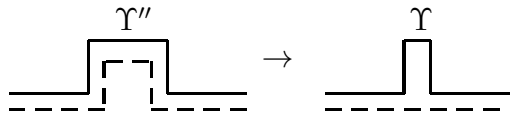
Fig. 3.} The different phases of the chiral symmetry in the states
of the hadronic bags of the $\Upsilon''$ and $\Upsilon$ levels.
Solid line is the hadron bag, dashed line is the quark condensate, that
partially penetrates into the $\Upsilon''$ bag and causes the additional
presure on the bag walls. This results in the less size of the
$\Upsilon''$ quarkonium in comparison with the purely Coulomb interaction.
In the case of the $\Upsilon$ quarkonium, the hadronic bag is small
and the condensate influence, taking place at large distances, is negligibly
small.
\end{description}
\begin{figure}[p]
\begin{center}
\begin{picture}(105,70)
\put(5,30){\line(1,0){15}}
\put(20,30){\line(0,1){10}}
\put(20,40){\line(1,0){15}}
\put(35,40){\line(0,-1){10}}
\put(35,30){\line(1,0){15}}

\put(25,42){$\Upsilon ''$}

\put(5,27){\line(1,0){3}}
\put(10,27){\line(1,0){3}}
\put(15,27){\line(1,0){3}}
\put(20,27){\line(1,0){3}}

\put(23,28){\line(0,1){3}}
\put(23,33){\line(0,1){3}}

\put(23.5,36){\line(1,0){3}}
\put(28.5,36){\line(1,0){3}}

\put(32,28){\line(0,1){3}}
\put(32,33){\line(0,1){3}}

\put(32,27){\line(1,0){3}}
\put(37,27){\line(1,0){3}}
\put(42,27){\line(1,0){3}}
\put(47,27){\line(1,0){3}}

\put(53,35){$\to$}

\put(65,30){\line(1,0){15}}
\put(80,30){\line(0,1){10}}
\put(80,40){\line(1,0){5}}
\put(85,40){\line(0,-1){10}}
\put(85,30){\line(1,0){15}}

\put(81,42){$\Upsilon$}

\put(65,27){\line(1,0){3}}
\put(70,27){\line(1,0){3}}
\put(75,27){\line(1,0){3}}
\put(80,27){\line(1,0){3}}
\put(85,27){\line(1,0){3}}
\put(90,27){\line(1,0){3}}
\put(95,27){\line(1,0){3}}

\end{picture}
\end{center}
Fig.3.
\end{figure}
\end{document}